\documentclass[twocolumn,prb,showpacs,floats,superscriptaddress]{revtex4}
\usepackage[dvips]{graphicx}
\usepackage{amsmath,bm,mathrsfs}
\usepackage{psfrag}

\begin{document}

\title{Averaged dynamics of two-phase media in a vibration field}

\author{Arthur V. Straube\footnote{E-mail: straube@stat.physik.uni-potsdam.de \\[0.5mm] {\mbox{} Paper published in Physics of Fluids {\bf{18}},
053303 (2006)}}} \affiliation{Department of Physics, University of
Potsdam, Am Neuen Palais 10, PF 601553, D-14415 Potsdam, Germany}
\author{Dmitry V. Lyubimov}
\affiliation{Theoretical Physics Department, Perm State
University, Bukirev 15, 614990 Perm, Russia}
\author{Sergey V. Shklyaev}
\affiliation{Theoretical Physics Department, Perm State
University, Bukirev 15, 614990 Perm, Russia}
\affiliation{Department of Mathematics, Technion--Israel Institute
of Technology, 32000 Haifa, Israel}

\date{\today}

\begin{abstract}
The averaged dynamics of various two-phase systems in a
high-frequency vibration field is studied theoretically. The
continuum approach is applied to describe such systems as solid
particle suspensions, emulsions, bubbly fluids, when the volume
concentration of the disperse phase is small and gravity is
insignificant. The dynamics of the disperse system is considered
by means of the method of averaging, when the fast pulsation and
slow averaged motion can be treated separately. Two averaged
models for both nondeformable and deformable particles, when the
compressibility of the disperse phase becomes important, are
obtained. A criterion when the compressibility of bubbles cannot
be neglected is figured out. For both cases the developed models
are applied to study the averaged dynamics of the disperse media
in an infinite plane layer under the action of transversal
vibration.
\end{abstract}


% 43. Acoustics
% 43.40.+s Structural acoustics and vibration
% 43.40.At Experimental and theoretical studies of vibrating systems

% 46. Continuum mechanics of solids
% 46.40.-f Vibrations and mechanical waves

% 47. Fluid Dynamics
% 47.55.-t Nonhomogeneous flows
% 47.55.Kf Multiphase and particle-laden flows
% 47.55.D- Drops and bubbles
% 47.15.Fe Stability of laminar flows
% 47.55.dd Bubble dynamics
% 47.35.-i Hydrodynamic waves

% 62. Mechanical and Acoustical Properties of Condensed Matter
% 62.30.+d Mechanical and elastic waves; vibrations

% 82. Physical Chemistry and Chemical Physics
% 82.70.-y Disperse systems; complex fluids

\maketitle

\section{Introduction}

Oscillatory processes can be found in a sheer uncountable number
of situations in nature at various time and length scales: from
subatomic to astronomic scales. Vibration is a mechanical
oscillatory process with an amplitude small compared to the
characteristic length scale of the system. Often, as in our study,
it is assumed that the inherent time scale of the system is much
larger than the period of the oscillation. Vibration mechanics has
been studied for a long time, general concepts of this nonlinear
phenomenon have been put into practice. It is widely used in
industry for transportation and separation of granular matter such
as powders and grains, to enhance mixing and reaction rates of
different species. In medicine, ultrasound is applied for
diagnostics and healing. In microfluidics and biotechnology it
serves as a means to manipulate physical (colloidal particles,
fine-dispersed powders, liquid drops in microemulsions) and
biological (cells, microorganisms, and micromolecules) objects
suspended in liquids. Quite a broad spectrum of applications has
motivated a great interest in the fundamental problems of
vibration dynamics. The present work is aimed at studying the
dynamics of two-phase media in a high-frequency vibration field.

There has been much effort to understand the impact of vibration
action on the behavior of systems of different nature.
Particularly, it has been understood that vibration can
substantially influence the dynamic and rheologic properties of
systems. An outstanding example is the Kapitza pendulum with a
periodically moving point of support, \cite{kapitza-51} where the
upper vertical state of the pendulum becomes stable. Other known
effects are a metal ball that lifts in ambient sand medium under
vibration, \cite{blekhman-00} granular matter that liquidizes
\cite{evesque-rajchenbach-89} or demonstrates the formation of
oscillons \cite{umbanhowar-etal-96} in a vibration field. As
typical hydrodynamic examples we mention excitation of a steady
relief at the oscillating interface pure liquid--particle
suspension, \cite{ivanova-etal-96, lyubimov-etal-99} and thermal
vibration convection, \cite{lyubimov-gersguni-98} where the action
of vibration on the nonuniform fluid results in generation of a
slow macroscopic motion even in the absence of gravity.

% {\bf [Below we briefly outline the principal results concerning
% the dynamics of a single inclusion in a fluid and collective
% behavior of particulate and bubbly media under vibration action.]}

The behavior of a single solid inclusion in fluid environment
under vibration has attracted much attention since long time ago.
As far back as 1831, Faraday used particles to visualize standing
acoustic waves and observed the effect of particle localization in
the acoustic field.\cite{faraday-1831} The dynamics of a solid
particle suspended in a vertically oscillating fluid was studied
in Ref.~\onlinecite{boyadzhiev-73}. It was shown that vibration
results in an average force, which can be used to affect particle
dynamics. Granat \cite{granat-60} considered theoretically a
motion of a solid sphere in a periodically oscillating flow of a
viscous fluid. For the fluid with the density higher than that of
the sphere, the latter oscillates with a smaller amplitude and a
lag in phase. Conversely, in a relatively less dense fluid the
oscillation of the sphere occurs with a larger amplitude with
respect to the flow and leads in phase. In an experimental study
by Chelomey,\cite{chelomey-83} the behavior of solid bodies in a
vertically vibrated vessel with a fluid was investigated. It was
observed that under vibration action, bodies with higher than
ambient density could rise to the surface, whereas light bodies
could sink. Theoretically, the averaged behavior of a solid sphere
suspended in oscillated fluid near a wall was discussed in
Ref.~\onlinecite{lamb-75} and has been recently
generalized.\cite{lugovtsov-sennitskii-01, lyubimov-etal-01} It
has been demonstrated that in addition to the gravity and
Archemedian forces there appears an averaged force that attracts
the body to the wall due to vibration. This vibration force can
lead to ascent of heavy or settling down of light bodies, as was
observed experimentally. \cite{chelomey-83}

The dynamics of a single compressible bubble in vibration fields
has been studied by many authors. The first analysis of the bubble
dynamics was made by Rayleigh \cite{rayleigh-17} (see also
Ref.~\onlinecite{lamb-75}). A nonlinear equation that couples the
radius of a bubble to the pressure in the liquid at large distance
from the bubble, known as the Rayleigh-Plesset or Rayleigh-Lamb
equation, was obtained and the eigenfrequencies of the small
oscillations were determined. Since then, many important aspects
of nonlinear bubble dynamics and cavitation hasve been studied,
e.g., natural and forced oscillations, parametric instability,
which causes the oscillation of the bubble shape against a
background of spherical pulsation, the influence of heat exchange
between bubbles and ambient fluid and nonideality of the gas
filling the bubble, the behavior of bubbles in nonuniform flows
(see reviews in Refs.~\onlinecite{plesset-prosperetti-77,
feng-leal-97}).

In the context of the time-averaged dynamics of a bubble an
important problem is obtaining the force exerted on the bubble due
to vibration. The behavior of a single bubble in a standing
acoustic wave is determined by ``the primary Bjerknes force''
after Bjerknes, \cite{bjerknes-06} who proposed its qualitative
explanation. The effect is quite general: the bubbles migrate to
the nodes of the pressure wave at high frequencies
$\omega>\omega_0$ and to antinodes at low frequencies
$\omega<\omega_0$, where $\omega_0$ is the frequency of natural
oscillation of the bubble. The average force on a particle, called
the force of radiation pressure, was obtained by Yosioka and
Kawasima.\cite{yosioka-kawasima-55} The considered setup comprised
a liquid drop in another liquid in a standing acoustic wave, where
both liquids were assumed to be compressible. A generalization for
the case of an arbitrary nonuniform oscillating flow has been
provided by Alekseev.\cite{alekseev-83} These results can be
applied to analyze both limiting cases -- the primary Bjerknes
effect for deformable bubbles and the behavior of a solid particle
in nonuniform pulsating flows.

Generalization of the description of a separate particle suspended
in a fluid to the case of a disperse medium (ensemble of many
particles) requires space averaging. The precise description of
all the disperse particles becomes redundant and unfeasible. There
are basically three possibilities to obtain a reasonable model:
(i) statistical approach, concerned with averaging over the
ensemble of particles; for solid particles and bubbles such a
procedure is carried out in
Refs.~\onlinecite{zhang-prosperetti-JFM-94,
zhang-prosperetti-PhF-94, bulthuis-etal-95}; (ii) derivation of a
kinetic equation (see, e.g., Refs.~\onlinecite{smereka-96,
russo-smereka-I-96, teshukov-gavrilyuk-02}); (iii) continuum
models, where after the space-averaging the disperse particles are
treated as a separate continuum.\cite{wijngaarden-72,
nigmatulin-91} In our study, we apply the last approach.

Although, wave propagation in disperse media is well understood,
the problem of the time-averaged dynamics for such systems has
received relatively little attention and remains rather
unexplored. Most of studies deal with acoustic vibration, when the
compressibility of carrier fluid is of crucial importance; the
feedback of particles on the flow is conventionally neglected. For
instance, Ganiev and Lapchinsky\cite{ganiev-lapchinsky-78} studied
averaged collective behavior of bubbles and solid particles
suspended in fluids in acoustic fields. The analog of the primary
Bjerknes effect was observed and simple explanation was provided.
In the present study, we concentrate on the averaged effects in
various disperse media under the action of high frequency, but
subacoustic vibration field, i.e., when the fluid is
incompressible. We also take into account the feedback of
particles on the carrier fluid and demonstrate that this is an
important factor, which can lead to nontrivial effects.

The paper is outlined as follows. In Sec.~\ref{sec:theor_model}
the theoretical model is developed for a suspension of
nondeformable particles. The equations describing the pulsation
and averaged dynamics are obtained. On the basis of this model the
stability of a quasiequilibrium state in a plane layer is analyzed
in Sec.~\ref{sec:stab_layer}. Section~\ref{sec:BF:theor_model}
addresses the vibrational dynamics of bubbly fluids. An averaged
model accounting for the compressibility of bubbles is obtained.
Particularly, it is shown that in the limit of weak
compressibility this model reduces to the model for nondeformable
particles, obtained in Sec.~\ref{sec:theor_model}. In
Sec.~\ref{sec:BF:layer}, the developed model is applied to
investigate the dynamics of a bubbly fluid in a plane layer under
vibration action. The results are summarized in
Sec.~\ref{sec:conclus}.

\section{Theoretical model of a nondeformable particle suspension in a vibration field \label{sec:theor_model}}
\subsection{Governing equations and basic assumptions}

Consider the behavior of an isothermal fluid (liquid or gas) laden
with disperse particles (solid particles, small droplets of
another liquid or bubbles) under the action of high frequency
vibration. We assume that all the particles are monodisperse
spheres of a radius $r_d$ and the amplitude of vibration $a$ is
small compared to this size: $a \ll r_d$ (for viscous pulsation a
milder restriction is enough to be held, see
Sec.~\ref{subsec:nondef_pulsations}).  The frequency of vibration
$\omega$ is assumed to be so high that the size of a viscous
boundary layer $\ell = \sqrt{\nu/\omega}$ ($\nu$ is the kinematic
viscosity of the fluid) near the walls of a container with the
suspension is small with respect to the characteristic scale $L$
of a flow. On the other hand, the scale $L$ is considered to be
small compared to the acoustic wave length $\lambda_a=2\pi
v_a/\omega$ ($v_a$ is the acoustic speed), which leads to a
twofold inequality $\nu/L^2 \ll \omega \ll v_a/L$.

We start with a two-fluid model valid for dilute suspensions, when
one can neglect interparticle interactions and interactions
between the particles and walls of a container. We assume that the
size $r_d$ is large enough to neglect diffusion of particles, but
small with respect to $L$. On a scale much greater than $r_d$ the
particles are regarded as a continuum with the volume fraction
$\varphi=4/3\pi r_d^{3} n$ (the volume fraction of the fluid phase
is $1-\varphi$), where $n$ is the number of particles per unit
volume of the medium. It is also supposed that the particles can
neither deform nor combine into agglomerates; the carrier and the
disperse phases have constant densities $\rho$ and $\rho_d$
(hereafter, the subscript ``$d$" stands for the disperse phase).
Since $\varphi$ differs from $n$ by a constant factor, in the
following $\varphi$ is called concentration. After averaging over
space the equations for mass and momentum of the phases
read:\cite{nigmatulin-91}
\vspace{-1mm}
\begin{subequations}
\label{massbal_eqs}
\begin{eqnarray}
{\rm div}\left( 1-\varphi \right){\bf U} +  {\rm div} \, \varphi
{\bf U}_d & = & 0, \\[-1mm]
\frac{d \varphi}{d t}+ \varphi \,{\rm div}\, {\bf U}_d & = & 0,
\label{massbal_eq_phi}
\end{eqnarray}
\end{subequations}
\vspace{-8mm}
\begin{subequations}
\label{momentum_eqs}
\begin{eqnarray}
\left(1-\varphi\right) \rho\frac{D \bf U}{D t} & = & - \nabla P +
\mu \nabla \cdot \left(1-\varphi \right){\bf \varepsilon}
-\varphi{\bf F}, \\
\rho_d\, \frac{d{\bf U}_d}{d t} & = & {\bf F},
\label{momentum_eq_part}
\end{eqnarray}
\end{subequations}
\noindent where ${\bf U}$ and ${\bf U}_d$ are the velocities of
the phases, $P$ is the pressure in fluid, $\mu$ is the fluid
viscosity, the shear rate tensor is given by
$\varepsilon_{ij}=\nabla_i U_j+\nabla_j U_i$. Derivatives
$D/Dt=\partial /\partial t+{\bf U} \cdot \nabla$ and
$d/dt=\partial /\partial t+{\bf U}_d \cdot \nabla$ are taken along
the Largangian path of the elements of the fluid and the disperse
phase, respectively.

The interphase interaction force can be written as
follows:\cite{nigmatulin-91, maxey-riley-83, yang-leal-91}
\begin{eqnarray} \label{iph_force}
{\bf F} & = & \rho \frac{D \bf U}{D t}
-\frac{9}{2r_d^2}\left(\frac{2/3+\kappa}{1+\kappa}\right){\bf W}
\nonumber \\[-1mm]
&&-\frac{9}{2r_d}\left(\frac{\kappa}{1+\kappa}\right)\sqrt{\frac{\rho\mu}{\pi}}\int\limits_0^t
\frac{d {\bf W}(\zeta)}{d \zeta}
\frac{d\zeta}{\sqrt{t-\zeta}}\nonumber \\[-1mm]
&&-\frac{1}{2}\rho\frac{d{\bf W}}{d t}-{\bf F}_m.
\end{eqnarray}
\noindent Here, the relative velocity of phases ${\bf W}={\bf
U}_d-{\bf U}$ and the ratio of viscosities of phases
$\kappa=\mu_d/\mu$ are introduced. The first term in
(\ref{iph_force}) is a force due to pressure gradient, the next
forces are the Stokes (Hadamard-Rybczynski) drag, \cite{lamb-75}
and three unsteady forces: the Basset history force, the added
mass force, allowing for the inertia of the surrounding fluid, and
a second memory force that is essential for non-solid particles.
\cite{yang-leal-91} We do not consider the effects that can be
initiated by rotation of particles, therefore the Magnus force is
not taken into account.

The relative contribution of the Stokes drag $F_S$, the Basset
$F_B$, and the added mass $F_M$ forces to the interphase force
(\ref{iph_force}) is governed by an ``unsteadiness parameter''
$K=r_d/ \sqrt{\nu t_*}$:
$$
\frac{F_B}{F_S} \sim K, \quad \frac{F_M}{F_S} \sim K^2,
$$
\noindent where $t_*$ is the timescale of the variation of the
relative velocity of phases. Note, that the pressure gradient term
and the added mass force in (\ref{iph_force}) are of the same
order. The same argument can be applied to the memory forces,
except for the limiting cases of an inviscid bubble ($\kappa=0$)
and a solid particle ($\kappa \gg 1$), when, either the Basset
term or the second memory force vanish, respectively.

For quasisteady flows ($K \ll 1$) and the flow regimes with finite
$K$, the result (\ref{iph_force}) holds provided that the particle
Reynolds number ${\rm Re}_d$ is small and gradients of the
velocity ${\bf U}$ are not too large: \cite{maxey-riley-83}
\begin{equation} \label{iph_force_cnds1}
{\rm Re}_d=\frac{r_d W}{\nu} \ll 1, \quad \frac{r_d^2 \,U}{\nu L}
\ll 1.
\end{equation}

\noindent In the case of substantially unsteady flows ($K\gg1$),
the force (\ref{iph_force}) has inviscid nature and the validity
criterion does not depend on the viscosity. The restrictions
(\ref{iph_force_cnds1}) should be replaced with the conditions
\begin{equation} \label{iph_force_cnds2}
\frac{W t_*}{r_d} \ll 1, \quad \frac{U t_*}{L} \ll 1,
\end{equation}
\noindent which arise from linearization requirements. Note, that
in this limit, $Re_d$ is not necessarily small and for finite and
large $Re_d$ the forces depending on viscosity in expression
(\ref{iph_force}) have to be modified. \cite{clift-etal-78}
However, for the flow with $K \gg 1$, these terms (irrespective of
the modification) are small compared to the leading pressure
gradient and added mass forces and therefore can be neglected.

The governing equations are formulated in an inertial reference
frame; the influence of vibration should be taken into account
through boundary conditions.

\subsection{Pulsation velocities} \label{subsec:nondef_pulsations}

The presence of two considerably different time scales, namely,
the characteristic oscillation time $1/\omega$ and the dissipative
time scale $L^2/\nu$, makes reasonable to use the multiscale
technique:\cite{nayfeh-81} the ``fast'' and the ``slow'' averaged
motions are treated separately. So, the operator of time
differentiation is replaced by a sum of two operators
$\partial/\partial t \to \omega \,\partial/\partial \tau +
\partial/\partial t$, where $\tau=\omega t$ is the fast time; the
actual fields of the velocities, the pressure, and the
concentration are represented by sums of the averaged (over the
fast time $\tau$) and the fast, or pulsation, components,
respectively: ${\bf U}={\bf u}+{\bf V}$, ${\bf U}_d={\bf u}_d+{\bf
V}_d$, $P=p+q$, $\varphi = c + \phi$.

Let us substitute this ansatz into the governing equations and
obtain the pulsation parts of the momentum Eqs.
(\ref{momentum_eqs}). We retain only the leading terms and take
into account a smallness of the concentration pulsation, $\phi
\sim (a/L)c$, which follows from Eq.~(\ref{massbal_eq_phi}). The
dimensionless variables are introduced using the following units:
$1/\omega$ and $L^2/\nu$ for the fast and the averaged time, $a
\omega$ and $\nu/L$ for the pulsation and the averaged velocities
of the phases, $\rho a \omega^2 L$ and $\rho \nu^2/L^2$ for the
pulsation and the averaged pressure, respectively. It is assumed
that the space derivatives can be evaluated with the scale $L$,
which means that the small scale dynamics in boundary layers near
solid walls and turbulent pulsations are not considered. As a
result, we arrive at the equations for the fast dynamics:
\begin{subequations}
\label{mass_puls}
\begin{eqnarray}
{\rm div}(1-c){\bf V} +  {\rm div}\,c{\bf V}_d & = & 0, \\
\frac{\partial \phi}{\partial \tau} + {\rm div}\,c{\bf V}_d & =
&0,
\end{eqnarray}
\end{subequations}
\vspace{-4.5mm}
\begin{subequations}
\label{momentum_puls}
\begin{eqnarray}
\frac{\partial {\bf V}}{\partial \tau} & = & - \nabla q + c \,{\bf
f}, \\
\delta \frac{\partial {\bf V}_d}{\partial \tau} & = &
\frac{\partial {\bf V}}{\partial \tau} - {\bf f},
\end{eqnarray}
\end{subequations}
\vspace{-3.5mm}
\begin{eqnarray}
{\bf f} &=&
\frac{1}{K^2}\left(\frac{2/3+\kappa}{1+\kappa}\right){\bf w} \nonumber \\
&&+\frac{1}{K\sqrt{\pi}}\left(\frac{\kappa}{1+\kappa}\right)
\int\limits_{0}^{\tau} \frac{\partial {\bf w}(\zeta)}{\partial
\zeta} \frac{d\zeta}{\sqrt{\tau-\zeta}} \nonumber \\
&& +\frac{1}{2}\frac{\partial {\bf w}}{\partial \tau}+{\bf f}_m,
\label{iphforce_puls}
\end{eqnarray}
\noindent where the relative pulsation velocity ${\bf w}={\bf V}_d
- {\bf V}$ and a dimensionless parameter $\delta=\rho_d/\rho$,
which is the ratio of densities of the phases, are introduced; the
pulsation component of the memory force ${\bf F}_m$ is denoted by
${\bf f}_m$. For the case of pulsation motion, the unsteadyness
parameter $K=r_d/\ell$ has the meaning of the ratio of the
particle radius to the size of a viscous boundary layer around it.

The restrictions for the amplitude of vibration $a$ are imposed by
the inequalities (\ref{iph_force_cnds1}), (\ref{iph_force_cnds2}),
and the necessity of linearization of the pulsation problem
(\ref{mass_puls})-(\ref{iphforce_puls}). For the motion with
finite $K$ and in the limit of inviscid pulsations, $K\gg1$, the
vibration amplitude is required to be small in the sense $a \ll
r_d$. However, in the limit of viscous pulsation, $K\ll 1$, a much
milder restriction for the vibration amplitude is enough to be
held: $a \ll \min \left( L, K^{-2}r_d\right)$.

Let us restrict our consideration to the case of the monochromatic
vibration and proceed to complex amplitudes: $\mathscr{F}({\bf
r},t)={\rm Re}\left[ \hat{\mathscr{F}}({\bf r}) \exp
(i\tau)\right]$. For such a vibration process, ${\bf \hat{f}}$
coincides (in dimensionless form) with the Fourier-transform of
the pulsation force ${\bf f}$ and can be represented as (cf.
Refs.~\onlinecite{yang-leal-91, kim-karrila-91}):
\begin{equation}
\hat{\bf f} = \chi \hat{\bf w}, \label{puls_force_f}
\end{equation}
\vspace{-5mm}
\begin{eqnarray}
&&\chi = \frac{9}{2K^2}\left(1+\lambda+\frac{\lambda^2}{9}\right.
\nonumber \\
&&\left.-\frac{(1+\lambda)^2f(\lambda)}{\kappa\left[\lambda^3-\lambda^2\tanh\lambda-2f(\lambda)\right]
+(\lambda+3)f(\lambda)} \right), \label{chi}
\end{eqnarray}
\noindent where $\lambda=(1+i)K/\sqrt{2}$ and
$f(\lambda)=\lambda^2\tanh\lambda-3\lambda+3\tanh\lambda$.

As follows from Eqs.~(\ref{mass_puls}), (\ref{momentum_puls}) and
relation (\ref{puls_force_f}), the pulsation velocities are not
independent, therefore the problem can be expressed in terms of an
auxiliary field ${\bf v}=(1-c){\bf \hat{V}}+c{\bf \hat{V}}_d$
having the meaning of the amplitude of the volume-weighted
velocity of the medium:
\begin{eqnarray}
& & \nabla  \times  S\, {\bf v} = 0, \quad {\rm div}\,{\bf v} = 0,
\label{puls_probl} \\[2mm]
& & S = \frac{\chi\rho_m+i\delta}{\chi+i\delta/\rho_e},
\label{S_expr}
\end{eqnarray}
\noindent where $\rho_m = 1 - c +c\delta$ is the density of the
medium and $1/\rho_e = 1-c+c/\delta$ is the specific volume. The
pulsation velocities expressed in terms of ${\bf v}$ are as
follows:
\begin{eqnarray}
\hat{\bf V} & = & \frac{\chi+i\delta}{\chi+i\delta/\rho_e}{\bf v}, \label{V_puls} \\[2mm]
\hat{\bf V}_d & = & \frac{\chi+i}{\chi+i\delta/\rho_e}{\bf v}.
\label{Vd_puls}
\end{eqnarray}
\noindent Note, that except for the limiting cases of viscous ($K
\ll 1$) and inviscid ($K \gg 1$) vibration regime, there appears a
phase shift in the pulsation velocities of the particles and
carrier fluid.

Now restrict our theory to the case of a closed nondeformable
cavity, filled entirely with the two-phase medium. Equations
(\ref{puls_probl}) should be supplemented with the boundary
conditions. The smallness of the vibration amplitude allows us to
prescribe the boundary conditions at the average position. Because
of incompressibility of the medium, the normal component of the
velocity ${\bf v}$ has to coincide with that of the boundary
velocity ${\bm \xi}$ at every point of the boundary $\Gamma$:
\begin{equation} \label{BC_puls}
{\bf n}\cdot{\bf v}|_{\Gamma}={\bf n}\cdot {\bm \xi},
\end{equation}
\noindent where ${\bf n}$ is the unit normal to the average
position of the boundary.

Assume that the system is subjected to translational, linearly
polarized vibration. In this case the boundary velocity of the
cavity is a constant vector: ${\bm \xi}={\bf j}$ (${\bf j}$ is the
unit vector aligned along the vibration axis).

Since the particle concentration is small, the solution to the
pulsation problem (\ref{puls_probl}), (\ref{BC_puls}) is sought as
a series in $c$: ${\bf v}={\bf v}_0+{\bf v}_1+\ldots$ with ${\bf
v}_1 \sim c$ (and analogously for $\hat{\bf V}$, $\hat{\bf V}_d$).
Then, to the leading order we obtain
\begin{equation} \label{V0}
{\bf v}_0=\hat{\bf V}_0={\bf j},
\end{equation}
i.e., the pulsation field is constant in the entire cavity:
uniform fluid moves together with the cavity as a solid. Hence, to
this approximation, the vibration does not influence the system,
therefore it is necessary to obtain a nonuniform correction to the
pulsation field.

Taking into account Eqs.~(\ref{puls_probl}), (\ref{BC_puls}) and
the result (\ref{V0}), to the next order we arrive at a problem
\begin{eqnarray}
\nabla \times {\bf v}_1 & = & S_1 \, {\bf j} \times \nabla c,
\quad {\rm div} \, {\bf v}_1=0, \label{V1_eqn} \\
{\bf n}\cdot{\bf v}_1|_{\Gamma} & = & 0, \nonumber \\
S_1 & = & \left(\delta-1\right) \frac{\chi+i}{\chi+i\delta},
\nonumber
\end{eqnarray}
\noindent where $S_1$ is a coefficient in a corresponding
expansion of (\ref{S_expr}): $S=1+S_1\, c+O(c^2)$. It becomes
clear from Eq.~(\ref{V1_eqn}), that a nonuniformity of the
pulsation field ${\bf v}_1$ is caused by interaction of vibration
with inhomogenieties in the particle concentration. We make a
note, that an explicit solution for the field ${\bf v}_1$ is not
needed while obtaining the averaged equations of motion: only the
vorticity and divergency of ${\bf v}_1$ are further used.

%only the {\bf solenoidal/curl} part of this field contributes to
%the averaged dynamics.

\subsection{Averaged equations. Single-fluid approximation}
\label{subsec:aver_motion}

We now turn to averaging of the governing equations. Let us write
out the averaged parts of Eqs.~(\ref{massbal_eqs}),
(\ref{momentum_eqs}) and retain only the leading terms. Note, that
averaged dynamics can be described in terms of a single-fluid
model. Contrastingly to the pulsation motion, relative
contributions of the Stokes drag, the memory forces, and the added
mass force to the averaged dynamics are defined by a small
parameter $r_d/L$; the unsteady forces turn out to be small
compared with the Stokes force and therefore can be neglected.
Because the difference in the averaged velocities of the phases
${\bf u}_d-{\bf u}$ is of order $r_d^2/L^2$, we neglect it
everywhere, except for the interphase force. In the latter, this
difference must be retained: despite its smallness, it is
multiplied by a vibration parameter
$\tilde{R}_v=a^2\omega^2L^2/\nu^2$, which can be large, and a
corresponding contribution in the averaged equations is generally
finite. As a result, we obtain the following equation of motion
\begin{eqnarray}
\rho_m\left( \frac{\partial {\bf u}}{\partial t} + {\bf u}\cdot
\nabla {\bf u}\right) &+& \tilde{R}_v \nabla \cdot \left[
\left(1-c \right)\overline{{\bf V}{\bf V}} + \delta c
\overline{{\bf
V}_d{\bf V}_d} \right] \nonumber \\
&= & -\nabla p + \nabla \cdot \left( 1-c \right)\varepsilon.
\end{eqnarray}
\noindent Here, $\varepsilon$ is the averaged shear rate tensor;
overlines are used to denote averaging over time. The accepted
approximation means, that the particles are ``frozen'' in the
averaged carrier flow.

It is important to note, that the nonuniformity of the pulsation
field can be caused not only by the considered mechanism, when
vibration interacts with nonuniformity of the medium due to
admixture. Another reason for a nonuniformity in the pulsation
field is interaction of vibration with the vorticity of the
averaged flow. A contribution to the averaged equations due to the
second mechanism is defined by the vector of pulsation transport
(see discussion in Ref.~\onlinecite{lyubimov-gersguni-98}).
However, for the linearly polarized vibration, the case under
consideration, this vector vanishes. A quite similar situation
takes place in thermal vibrational
convection.\cite{lyubimov-gersguni-98} The only difference is that
there the nonuniformity of density is caused by nonisothermality
of fluid, but not due to the presence of a disperse phase.

Using (\ref{V_puls}), (\ref{Vd_puls}), we calculate the averaged
Reynolds stress
\begin{eqnarray}
&&\left(1-c \right)\overline{{\bf V}{\bf V}} + \delta c
\overline{{\bf V}_d{\bf V}_d} = G \,{\rm Re}\left( {\bf v}{\bf
v}^{\ast}\right), \\
G & = &
\frac{1}{2}(1-c)\left|\frac{\chi+i\delta}{\chi+i\delta/\rho_e}\right|^2
+ \frac{1}{2}\delta c
\left|\frac{\chi+i}{\chi+i\delta/\rho_e}\right|^2.
\end{eqnarray}
Taking into account smallness of the particle concentration, the
result (\ref{V0}), and Eqs. (\ref{puls_probl}), we evaluate the
``vibration force''
\begin{eqnarray}
\tilde{R}_v \nabla \cdot \left[ G\,{\rm Re} \left( {\bf v}{\bf
v}^{\ast} \right) \right] & = & -\tilde{R}_v B\, {\bf j}\,{\bf
j}\cdot \nabla c + \ldots, \nonumber \\
B & = & \frac{\left(\delta-1\right)^2{\rm Im}\chi}{2\left|
\chi+i\delta \right|^2}, \label{param_B}
\end{eqnarray}
\noindent where dots denote the omitted gradient terms. Such terms
lead only to a renormalization of the pressure, which is not
important in the framework of our consideration.

In the vein of the Boussinesq approximation, we neglect the
nonuniformities of the density everywhere except for the the
vibration force, where the small concentration is multiplied by
the vibration parameter $\tilde{R}_v$, which is generally large.
Such a transformation corresponds to a limiting transition, when
$\tilde{R}_v\to\infty$, $c \to 0$, but their product $\tilde{R}_v
c_{\ast}$ is finite (here $c_{\ast}$ is the reference value of the
particle concentration).

Averaging of the mass balance equations is straightforward.
Finally, we end up with a set of the averaged equations describing
the behavior of a suspension of nondeformable particles in a high
frequency vibration field:
\begin{eqnarray}
\frac{\partial {\bf u}}{\partial t}+{\bf u}\cdot\nabla {\bf u}-R_v
\, {\bf j}\,{\bf j}\cdot \nabla c & = & -\nabla \Pi
+\nabla^2{\bf u}, \label{aveq_vel}\\
\frac{\partial c}{\partial t} + {\bf u}\cdot\nabla c & = & 0,
\quad {\rm div}\,{\bf u}=0, \label{aveq_conc}
\end{eqnarray}
\noindent where $\Pi$ is the renormalized pressure, the
concentration is normalized by $c_{\ast}$, and the vibration
parameter is redefined $R_v=a^2\omega^2 L^2 B c_{\ast}/\nu^2$.

The vibration force, appeared as a result of averaging of the
equations, is caused by interaction of the monochromatic linearly
polarized vibration with the inhomogeneity of medium due to a
nonuniform distribution of particles; for the uniform distribution
of particles it vanishes. The force is directed along the
vibration axis and takes on maximal values in the domains of the
flow where the gradient of particle concentration coincides with
the direction of the vibration axis. For the initial state with a
nonuniform distribution of particles and the vanishing averaged
velocity the action of the vibration force can result in
generation of averaged motion.

It is important to make a note of the generic character of the
vibration force. The dependence on the parameters $\delta$,
$\kappa$, $K$, i.e., on the the nature of the particles, is
gathered in the parameter $B$. This allows us to make two
conclusions. First, since $B>0$ for all values of these
parameters, the direction of the force is determined only by the
concentration field and an orientation of the vibration axis.
Second, for different media only the intensity of response towards
vibration action changes. However, such a quantitative distinction
is worth discussing for comparison with experiments, therefore
properties of the function $B=B(\delta, \kappa, K)$ are discussed
in Appendix.

The set of averaged Eqs.~(\ref{aveq_vel}), (\ref{aveq_conc}) must
be supplemented with boundary conditions for the velocity. For the
considered case of vibration action, the Schlichting mechanism
\cite{schlichting-68} of surface generation of averaged flow is
absent, therefore on a rigid surface $\Gamma$ one should use the
conventional no-slip boundary condition
\begin{equation}
{\bf u}|_{\Gamma}=0. \label{aveBC}
\end{equation}

Indeed, the intensity of the averaged flow caused by the
Schlichting mechanism is governed by the pulsation Reynolds number
\cite{lyubimov-gersguni-98} ${\rm Re}_p=a^2\omega/\nu$. Compared
to the considered bulk mechanism, which is governed by the
parameter $R_v$, the Schlichting mechanism can give a finite
contribution to generation of averaged motion only if the leading
part of the pulsation velocity is nonuniform. In our case, this
requirement is not satisfied: according to (\ref{V0}), the leading
part of the pulsation velocity is uniform, a nonuniformity occurs
only as a small correction to the uniform term. Hence, the
Schlichting generation of averaged flow is negligible compared to
the contribution of the bulk mechanism. Rather similar situation
takes place in thermal vibration convection: for the same kind of
vibration action on weakly nonuniform fluid, when the density
nonuniformity is caused by nonisothermality of fluid, the
Schlichting mechanism does not contribute to generation of average
flow.

\section{Stability of a quasiequilibrium in vibration field}
\label{sec:stab_layer}
\subsection{Basic quasiequilibrium state}

Let a suspension fill an infinite plane layer $-1 < z < 1$ (of a
dimensional width $2h$) in weightlessness. Consider the influence
of the transversal linearly polarized vibration on the stability
of a quasiequilibrium. A quasiequilibrium is thought of as a state
in which the averaged velocity vanishes in the entire volume of
the system, whereas the pulsation velocity is different from zero.

This problem is described by Eqs.~(\ref{aveq_vel}),
(\ref{aveq_conc}) with ${\bf j}=(0,0,1)$ and boundary conditions
(\ref{aveBC}). By setting ${\bf u}=0$ in the equations, we obtain
the conditions defining the quasiequilibrium state
\begin{equation}
R_v{\bf j\,j}\cdot\nabla c_0 = \nabla \Pi_0, \quad c_0={\rm
const}(t), \nonumber
\end{equation}
\noindent where the subscript ``0'' indicates the state of
quasiequilibrium. Integration of the first equation leads to a
condition
\begin{equation} \label{quasi-eq_cnd}
R_v c_0(x,y,z) = \Pi_0(z)+f(x,y).
\end{equation}
\noindent Here $f(x,y)$ is an arbitrary function of longitudinal
coordinates. Thus, there is a set of possible quasiequilibrium
distributions of particles that is defined by a sum of a function
of only transversal coordinate and a function of longitudinal
coordinates.

As a quasiequilibrium concentration we choose a linear function of
the transversal coordinate
\begin{equation} \label{c0}
c_0 = \alpha + \beta z, \quad \alpha>0, \quad |\beta|<\alpha
\end{equation}
\noindent and investigate its stability. Although gravity is
unimportant in the framework of our theory, in a quiescent fluid
(in the absence of vibration) it ultimately leads to
stratification of particles. For this reason, the choice of the
initial distribution of particles in the form (\ref{c0}) is rather
natural for a narrow layer. Further, as a reference value for the
particle concentration we use $\beta h$, therefore $R_v=a^2
\omega^2 h^3 B \beta/\nu^2$.

\subsection{Statement of linear stability problem}

To investigate the stability of the quasiequilibrium state we
introduce perturbations of the velocity ${\bf u}$, the pressure
$q$, and the concentration $\phi$ , and substitute the disturbed
fields ${\bf u}$, $\Pi_0+q$, $c_0+\phi$ into the problem
(\ref{aveq_vel})-(\ref{aveBC}). We linearize the equations near
the quasiequilibrium and obtain a set of equations and boundary
conditions for small perturbations
\begin{eqnarray}
\frac{\partial {\bf u}}{\partial t} -R_v \, {\bf j}\,{\bf j}
\cdot \nabla \phi & = & -\nabla q +\nabla^2 {\bf u}, \nonumber \\
\frac{\partial \phi}{\partial t} + {\bf u} \cdot \nabla c_0 & = &
0,
\quad {\rm div} \, {\bf u}=0, \nonumber \\
z & = & \pm 1: \; {\bf u}=0. \nonumber
\end{eqnarray}

Since the quasiequilibrium state is uniform and isotropic in a
plane of the layer, we can restrict our stability analysis to the
case of the two-dimensional perturbations in the form of rolls.
Assuming all the perturbations to be proportional to $\exp(\lambda
t - iky)$, where $\lambda=\lambda_r+i\lambda_i$ is the complex
growth rate and $k$ is the real wave number, we rewrite the
problem for one variable. Eliminating the pressure from the
equation of motion and taking into account the continuity
equation, we obtain for the amplitude $\hat{\phi}$ of the
concentration perturbation (the perturbations of the concentration
and the transversal component of the velocity differ only by a
constant factor):
\begin{eqnarray}
\lambda^2 D^2 \hat{\phi}-R_v k^2 \hat{\phi}^{\prime} =
\lambda D^4 \hat{\phi}, \label{eq_phi} \\
z=\pm 1: \; \hat{\phi}=0, \; \hat{\phi}^{\prime}=0, \label{bc_phi}
\end{eqnarray}
\noindent where a differential operator $D^2=d^2/dz^2-k^2$ is
introduced. While obtaining (\ref{eq_phi}), (\ref{bc_phi}) it was
assumed that $\lambda \ne 0$. In the opposite case, $\lambda=0$,
there exist obvious solutions $\hat{\phi}={\rm const}$ for $k \ne
0$ and $\hat{\phi}=\hat{\phi}(z)$ for $kR_v=0$; the perturbations
of the velocity vanish for both these cases. The existence of
these neutrally stable solutions is easy to explain: although the
mentioned perturbations deform the concentration field (\ref{c0}),
it still belongs to the set of quasiequilibrium solutions
(\ref{quasi-eq_cnd}).

The boundary value problem (\ref{eq_phi}), (\ref{bc_phi}) is a
spectral amplitude problem, the condition of nontrivial solution
defines the growth rate $\lambda$ as a function of parameters
$R_v$, $k$. In the absence of vibration ($R_v=0$), besides the
discussed solutions for $\lambda=0$, the problem admits an
explicit solution, \cite{gershuni-zhuhovitsky-76} eigenvalues of
which belong to a spectrum of real negative values: all the
perturbations decay and the system is stable. Further we are
interested in the nontrivial case of $R_v \ne 0$. In this case the
problem (\ref{eq_phi}), (\ref{bc_phi}) is no longer self-adjoint,
therefore the spectrum of growth rates can be complex.

Because the parameter $\beta$ is not of the fixed sign, the
vibration parameter can take on both positive and negative values.
However, the problem is invariant with respect to the mutual
change of sign of the transversal coordinate and the vibration
parameter. Such a symmetry is caused by the absence of gravity,
which allows us to restrict our analysis to the case $R_v>0$.

\subsection{The case of small values of vibration number $R_v$}

Consider the case of small intensities of vibration action. We
look for a solution to the problem (\ref{eq_phi}), (\ref{bc_phi})
in the form of series in small $R_v$
\begin{equation}
\hat{\phi}=\hat{\phi}_0+R_v \hat{\phi}_1+R_v^2
\hat{\phi}_2+\ldots, \quad \lambda=\lambda_0+R_v \lambda_1+R_v^2
\lambda_2+\ldots.
\end{equation}

Substituting these expansions into Eqs.~(\ref{eq_phi}),
(\ref{bc_phi}), we find that $\hat{\phi}_0$ and $\lambda_1$ are
defined by the following problem
\begin{eqnarray}
\lambda_1 D^4 \hat{\phi}_0+k^2\hat{\phi}'_0 = 0, \label{eq_phi0} \\
\hat{\phi}_0(\pm 1)=0, \; \hat{\phi}_0^{\prime}(\pm 1)=0.
\label{bc_phi0}
\end{eqnarray}

Let us multiply Eq.~(\ref{eq_phi0}) by the complex conjugate
$\hat{\phi}_0^{\ast}$ and integrate by parts across the layer.
Taking into account (\ref{bc_phi0}), after straightforward
transformation we recognize that $\lambda_1=i\omega_1$, where
$\omega_1$ is a real number. To the second order we obtain the
problem
\begin{eqnarray}
\lambda_1 D^4 \hat{\phi}_1+k^2\hat{\phi}'_1 = \lambda_1^2
D^2\hat{\phi}_0-\lambda_2 D^4\hat{\phi}_0, \label{eq_phi1} \\
\hat{\phi}_1(\pm 1)=0, \; \hat{\phi}_1^{\prime}(\pm 1)=0.
\label{bc_phi1}
\end{eqnarray}
Let us formulate the solvability condition for problem
(\ref{eq_phi1}), (\ref{bc_phi1}). We multiply Eq.~(\ref{eq_phi1})
by $\hat{\phi}_0^{\ast}$, the solution of the problem adjoint to
(\ref{eq_phi}), (\ref{bc_phi}), and integrate across the layer.
Taking into account boundary conditions (\ref{bc_phi0}),
(\ref{bc_phi1}), we end up with
\begin{equation}
\lambda_2=\frac{\omega_1^2 \left<
|\hat{\phi}_0^{\prime}|^2+k^2|\hat{\phi}_0|^2
\right>}{\left<|\hat{\phi}_0^{\prime\prime}|^2+2k^2|\hat{\phi}_0^{\prime}|^2+k^4|\hat{\phi}_0|^2\right>}>0,
\end{equation}
\noindent where $\left<...\right>=\int_{-1}^{1}...\,dz$.

Thus, for any value of the wave number $k$ the growth rate
$\lambda_2$ is real and positive, and hence for any arbitrarily
small value of the vibration parameter $R_v$ there appears
oscillatory instability.
\begin{figure}[!b]
\vspace{-2mm}
\includegraphics[width=5.0cm]{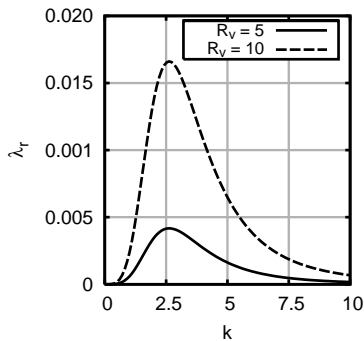}
\caption{The real part of the growth rate $\lambda_r$ as a
function of the wave number $k$.} \label{fig:1}
\end{figure}

\subsection{The case of finite values of vibration number $R_v$}

\begin{figure}[!h]
\vspace{-2.5mm}
\includegraphics[width=5.0cm]{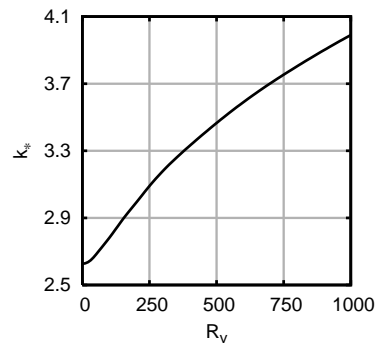}
\caption{The critical value $k_{\ast}$ of the wave number versus
vibration parameter $R_v$.} \label{fig:2}
\end{figure}

In the case of arbitrary values of the vibration parameter $R_v$
the boundary value problem (\ref{eq_phi}), (\ref{bc_phi}) was
solved numerically. First, the spectrum of eigenvalues was
analyzed by the Galerkin method with a basis given by the
eigenfunctions \cite{gershuni-zhuhovitsky-76} of the problem
(\ref{eq_phi}), (\ref{bc_phi}) at $R_v=0$, $\lambda \ne 0$. Next,
the behavior of the mode with the largest $\lambda_r$ was
specified by the standard shooting method.\cite{stoer-bulirsch-80}
It was found that in all the range of the parameters $k$ and
$R_v>0$ there exists a growth rate with the positive real part
$\lambda_r>0$: the system turns out to be unstable for any
intensity of vibration action. The real part of the growth rate as
a function of the wave number (the branch with the maximal
$\lambda_r$) is plotted in Fig.~\ref{fig:1} for different values
of vibration parameter $R_v$. For any value of the vibration
parameter $R_v$ the real part of the growth rate $\lambda_r$ has a
maximum at some value $k_{\ast}$ of the wave number. This critical
value $k_{\ast}$ defines the perturbation with the fastest growth
rate, which dependence on the vibration parameter is given in
Fig.~\ref{fig:2}. At small values of $R_v$ the curve is outgoing
from a point with a finite value of $k_{\ast}$. With the growth of
vibration intensity the characteristic length scale
$k_{\ast}^{-1}$ of hydrodynamic patterns developing as a result of
instability of a quasiequilibrium state monotonically decreases.

\begin{figure}[!h]
\vspace{-2.5mm}
\includegraphics[width=4.7cm]{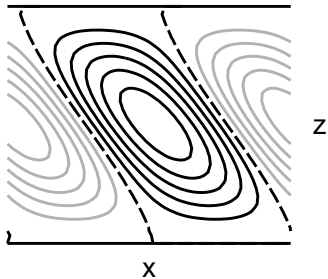}
\caption{Streamlines of the averaged flow for the perturbation
with the fastest growth rate, $R_v=10$.} \label{fig:3}
\end{figure}

Figure~\ref{fig:3} demonstrates the eigenfunction $\hat{\phi}$,
which corresponds to the breakdown of the quasiequilibrium and
differs from the transversal velocity component by a constant
factor. Under vibration action a nonuniformity of the
quasiequilibrium distribution of particles leads to generation of
the averaged fluid motion, which, in its turn, deforms the initial
distribution of particles.
%

%%%%%%%%%%%%%%%%%%%%%%%%%%%%%%%%%%%%%%%%%%%%%%%%%%%%%%%%%%%%%%%%%%%%%%%%%%%%%%%

\section{Averaged dynamics of a bubbly fluid} \label{sec:BF:theor_model}
\subsection{Basic equations}

So far, the disperse phase was assumed to be nondeformable.
However, the description of a bubbly fluid in vibration field
requires a closer look. The presence of the disperse phase that is
able to change its volume under the action of an external periodic
field, can drastically influence the pulsation and the averaged
dynamics of the system. Next, the case of a monodisperse bubbly
fluid is studied and the criterion when compressibility of the
disperse phase cannot be neglected is provided. Assuming the
concentration $\varphi$ of bubbles to be small, we formulate the
equations for the momentum of the liquid carrier phase and the
bubbles (\ref{momentum_eqs}) together with (\ref{iph_force}),
where for our particular consideration $\kappa=0$. An essential
modification that has to be taken into account here concerns an
additional contribution to the interphase force
\cite{nigmatulin-91}
\begin{equation}
{\bf F}_c=-\rho\frac{3b}{2r_d}{\bf W}, \nonumber
\end{equation}
\noindent where $b$ is the space-averaged radial velocity of
bubbles.

Further, assuming that the size of viscous boundary layer $\ell$
is small compared with the time-averaged radius $R$ of a bubble
$(\ell \ll R)$, we can neglect the memory forces.
\cite{yang-leal-91} Indeed, in the frames of the accepted
approximation the leading contribution to the pulsation motion is
made by the inertia forces [the first and the fourth terms in
(\ref{iph_force})], whereas for the averaged motion the Stokes
force and ${\bf F}_c$ are dominant. We suppose the density of the
gaseous phase to be much smaller than the density of the fluid and
neglect the left-hand side in Eq.~(\ref{momentum_eq_part}), i.e.,
the force on the bubbles vanishes. The equations defining the
kinematics read \cite{nigmatulin-91}
\begin{eqnarray}
{\rm div}\left( 1-\varphi \right){\bf U} +  {\rm div} \, \varphi
{\bf U}_d & = & \frac{3\varphi}{r_d}\,b, \label{BF:mass_eq_fluid}\\
\frac{d \varphi}{d t}+ \varphi \,{\rm div}\, {\bf U}_d & = &
\frac{3\varphi}{r_d}\,b, \label{BF:mass_eq_bubbles}
\end{eqnarray}
\begin{equation}
\frac{d \rho_d}{d t}=-\frac{3}{r_d}\,b\rho_d, \quad
\frac{dr_d}{dt}=b.
\end{equation}
The pressure in the fluid $P$, the pressure in a bubble $P_g$ and
the velocity of the radial oscillation of a bubble are coupled
through the Rayleigh-Plesset equation:\cite{nigmatulin-91,
brennen-95}
\begin{equation}
r_d\frac{db}{dt}+\frac{3}{2}b^2-\frac{1}{4}{\bf
W}^2=\frac{1}{\rho}\left(P_g-\frac{2\sigma}{r_d}-P\right).
\end{equation}
\noindent where $\sigma$ is the surface tension.

According to the procedure laid down in
Sec.~\ref{sec:theor_model}, we introduce the time hierarchy,
splitting all the fields into the pulsation (depending on the fast
time $\tau$) and the averaged parts. The notation and units for
the pulsation and averaged components of the velocities, the
pressure, and the concentration are the same as used before, in
Sec.~\ref{sec:theor_model}. Hereafter, $b$ is the pulsation
velocity of radial motion.

\subsection{Pulsation motion}\label{subsec:BF:pulsations}

Consider vibration of a cavity entirely filled with a bubbly
fluid. The principal assumptions introduced in
Sec.~\ref{sec:theor_model} are kept the same: the carrier phase is
incompressible, the frequency of vibration is high, but
subacoustic, the size of bubbles and the vibration amplitude are
small. On the other hand, we impose stronger restrictions for the
amplitude $a$ and the frequency $\omega$ of vibration:
\begin{equation}
aL \ll R^2, \quad \omega \gg \frac{{\rm max}(\nu,
\chi,\chi_d)}{R^2},
\end{equation}
\noindent where $\chi$ and $\chi_d$ are the thermal diffusivities
of the phases. The second inequality implies that dissipation is
unimportant for the pulsation motion, therefore the pressure
oscillation of gas obeys an adiabatic law.

Let us now formulate the equations for the pulsation motion. The
units for the pulsation velocity and the pressure are kept the
same as earlier (see Sec.~\ref{subsec:nondef_pulsations}), as a
unit for the velocity of the radial motion we choose $a\omega
L/R$. The equation for the pulsation of the concentration $\phi$
splits off and can be treated separately. For this field we obtain
\begin{equation} \label{BF:conc_puls}
\frac{\partial \phi}{\partial
\tau}=3\,\frac{aL}{R^2}\left(bc-\frac{1}{l}\,{\rm div}\,c{\bf
V}\right), \quad l=\frac{L}{R} \gg 1.
\end{equation}
\noindent As it follows from Eq.~(\ref{BF:conc_puls}), the
pulsation of the concentration is small compared to the averaged
field $c$. Note, that the second term in the right-hand side of
Eq.~(\ref{BF:conc_puls}) is, generally, much smaller than the
first one. However, under certain conditions it becomes
non-negligible (see Sec.~\ref{subsec:low_freq}).

It is convenient to write down the equations for the remaining
pulsation fields in terms of amplitudes. Assuming that all the
fields are proportional to $\exp i\tau$, we obtain
\begin{eqnarray}
i(1-c)\hat{\bf V}=-\nabla\hat{q}, \quad \hat{\bf V}_d=3\hat{\bf
V}, \label{BF:momentum_puls} \\
{\rm div}\,(1+2c)\hat{\bf V}=3\,cl^2\hat{b}, \quad
\hat{q}=i\hat{b}(\Omega^2-1), \label{BF:mass_puls}
\end{eqnarray}
\noindent where $\hat{\bf V}$, $\hat{\bf V}_d$, $\hat{q}$,
$\hat{b}$ are the complex amplitudes of the velocities of the
phases, the pressure, and the velocity of the radial motion.
Equations (\ref{BF:momentum_puls}), (\ref{BF:mass_puls}) contain
the following dimensionless parameter:
\begin{equation}
\Omega^2=\frac{1}{\rho \omega^2 R^2} \left( 3\gamma P_g -
\frac{2\sigma}{R} \right).
\end{equation}
\noindent Here, $\Omega$ is the ratio of the eigenfrequency of the
volume oscillation of a bubble to the frequency of vibration
$\omega$, the parameter $\gamma$ is the adiabatic exponent, and
$P_g$ is the averaged pressure inside a bubble. Note the known
relation between the velocities of the phases in a vibration
field:\cite{landau-lifshitz-87} the amplitude of oscillation of a
bubble in an inviscid fluid is three times as much as that of the
fluid.

Equation~(\ref{BF:mass_puls}) contains a product of two asymptotic
parameters: large $l^2$ and small $c$. Depending on the relation
between these parameters, different solutions to the pulsation
problem is possible. Let us introduce a renormalized concentration
$\Phi({\bf r})=l^2c\,({\bf r})$. Assuming that this field is
finite,\cite{caflisch-etal-85} we obtain a Helmholtz equation for
a potential $\psi$ of the fluid velocity and the impermeability
boundary condition:
\begin{eqnarray}
\nabla^2 \psi+\frac{3\Phi({\bf r})}{\Omega^2-1}\psi & = & 0, \label{BF:eq_Phi_gen}\\
\nabla_n \psi |_{\Gamma} & = & j_n.  \label{BF:bc_Phi_gen}
\end{eqnarray}
The amplitudes of the pulsation fields are expressed in terms of
the potential $\psi$ as follows:
\begin{eqnarray}\label{BF:ampl}
\hat{\bf V}=\nabla\psi, \quad \hat{b}=-\frac{\psi}{\Omega^2-1},
\quad \hat{q}=-i\psi.
\end{eqnarray}

Equation (\ref{BF:eq_Phi_gen}) has a form of the stationary
Schr\"odinger equation, in which the role of the scattering
potential is played by the averaged concentration field $\Phi({\bf
r})$ -- the waves are scattered on the nonuniformities of the
bubble distribution. The impermeability condition
(\ref{BF:bc_Phi_gen}) should be imposed on the volume-weighted
velocity $(1-c)\hat{\bf V}+ c\hat{\bf V}_d$, which by virtue of
(\ref{BF:momentum_puls}) differs from $\hat{\bf V}$ only by a
small term $2c\hat{\bf V}$. Note that as the frequency $\Omega$
gets closer to a resonant value $\Omega=1$, the susceptibility of
the system to vibration action gets much higher. At the point of
the resonance, the performed analysis is no longer valid, because
even weak dissipation becomes important and cannot be neglected.

Generally, Eq.~(\ref{BF:eq_Phi_gen}) can be solved only
numerically. However, in the case of a small parameter
$\epsilon=\left<\Phi\right>/(\Omega^2-1)$, it is possible to
obtain the leading part of the solution. Here, $\left<\Phi\right>$
is the value of $\Phi$ averaged all over the volume of the system.
Note that the parameter $\epsilon$ can be small for two different
reasons: for small concentration $\left<\Phi\right>$ or for small
vibration frequencies, $\Omega \gg 1$. Assuming that $z$-axis is
aligned along the direction of vibration, we can represent the
solution in the form
\begin{equation}
\psi=z-z_c+\psi_1, \nonumber
\end{equation}
\noindent where the term $z-z_c$ corresponds to the solution of
(\ref{BF:eq_Phi_gen}), (\ref{BF:bc_Phi_gen}) for $\Phi({\bf r})=0$
and $\psi_1\sim\epsilon$ is defined by a boundary value problem
\begin{eqnarray}
\nabla^2\psi_1 & = & -\frac{3\Phi({\bf r})}{\Omega^2-1}(z-z_c), \label{BF:eq_psi1} \\
\nabla_n\psi_1 |_{\Gamma} & = & 0. \label{BF:bc_psi1}
\end{eqnarray}
Applying the solvability conditions for the problem
(\ref{BF:eq_psi1}), (\ref{BF:bc_psi1}), we find the constant
\begin{equation} \label{BF:z_c}
z_c=\frac{\int_V\Phi z \, dV}{\int_V\Phi \, dV},
\end{equation}
\noindent which has the meaning of the $z$-coordinate of the
center of mass for the distribution of bubbles.

Thus, for the case of small $\epsilon$ we obtain
\begin{equation}\label{BF:V}
\hat{\bf V}={\bf j}+O(\epsilon), \quad
\hat{b}=-\frac{z-z_c}{\Omega^2-1}+O(\epsilon).
\end{equation}

\subsection{Averaged motion}\label{subsec:BF:aver_motion}

We now proceed to the averaging of the governing equations. The
equation of the momentum for the averaged motion of the fluid
takes the following form:
\begin{equation} \label{BF:eq_u_averaging}
(1-c)\left(\frac{D {\bf u}}{D t}+ \tilde{R}_v \overline{{\bf
V}\cdot\nabla{\bf V}} \right)-\tilde{R}_v\frac{L}{a}
\,\overline{\phi\frac{\partial{\bf V}}{\partial \tau}}=-\nabla
p+\nabla\cdot(1-c)\varepsilon.
\end{equation}
\noindent Here, as in Sec.~\ref{sec:theor_model},
$\tilde{R}_v=a^2\omega^2L^2/\nu^2$ and $\varepsilon$ is the
averaged shear rate tensor. From the averaged equation for the
bubbles follows
\begin{equation}
\frac{2}{3} \frac{r_d^2}{L^2} \tilde{R}_v\, \overline{{\bf
V}\cdot\nabla{\bf V}} + \tilde{R}_v \, \overline{b{\bf V}} + {\bf
u}_d - {\bf u} =0, \nonumber
\end{equation}
\noindent where the first term is small compared to the others.
With account of (\ref{BF:ampl}), this equation provides a relation
between the averaged velocities of the phases:
\begin{equation} \label{BF:u_d}
{\bf u}_d={\bf u}+Q\nabla \psi^2, \quad
Q=\frac{1}{4}\,\frac{\tilde{R}_v}{\Omega^2-1}.
\end{equation}
\noindent Here $Q$ is the vibration parameter, which is different
to that introduced in Sec.~\ref{sec:theor_model}.

To perform averaging in Eq.~(\ref{BF:eq_u_averaging}) we take into
account relations (\ref{BF:ampl}) and Eq.~(\ref{BF:conc_puls}).
Retaining only the leading terms, we arrive at the equation
\begin{equation} \label{BF:eq_u}
\frac{\partial {\bf u}}{\partial t}+{\bf u}\cdot\nabla{\bf
u}=-\nabla \Pi +\nabla^2{\bf u}+ 3Q \Phi\nabla \psi^2.
\end{equation}
While obtaining this equation, we neglect the terms proportional
to $c$ and renormalize the pressure; the renormalized variable is
denoted by $\Pi$. Variations of the averaged pressure, having in
dimensional units the order $\rho b^2 \sim \rho (a\omega l)^2$,
are assumed to be much smaller than the averaged pressure $P_g$
inside a bubble. This assumption ensures that the averaged radius
of the bubbles keeps constant and equals $R$. The absolute value
of the averaged pressure becomes insignificant, only its gradient
is of importance.

The averaging of the equation for the bubble concentration
(\ref{BF:mass_eq_bubbles}) is straightforward and results in
\begin{equation} \label{BF:eq_c}
\frac{\partial \Phi}{\partial t}+{\rm div}\,\Phi {\bf u}_d=0.
\end{equation}
\noindent The right-hand side of Eq.~(\ref{BF:eq_c}) vanishes,
because the phases of $b$ and ${\bf V}$ are shifted with respect
to $\phi$ and $r_d$ by a quarter-period. For the same reason,
there is no nontrivial contribution to the averaged mass balance
equation, which transforms to the incompressibility condition for
the fluid
\begin{equation} \label{BF:eq_incompr}
{\rm div}\, {\bf u}=0.
\end{equation}

Equations (\ref{BF:u_d})-(\ref{BF:eq_incompr}) together with
Eqs.~(\ref{BF:eq_Phi_gen}), (\ref{BF:bc_Phi_gen}) defining the
pulsation potential $\psi$, represent a set of equations for the
averaged dynamics of a bubbly fluid.

In perfect analogy with the case of nondeformable particles, the
averaged motion induced in a boundary layer is negligibly small;
the principal effect is caused by the considered mechanism of bulk
generation. Thus, at rigid boundaries the conventional no-slip
boundary condition for the averaged fluid velocity should be
imposed. According to (\ref{BF:u_d}), the velocity of the bubbles
at the boundaries does not turn to zero. Thus, the boundary
conditions for the concentration of bubbles must be prescribed at
the inflow boundaries, i.e., where ${\bf n}\cdot{\bf u}_d>0$
(${\bf n}$ is the outward normal to a boundary).

Note that the intensity of the averaged flow of the bubbly fluid
is significantly (by a factor of $l^2$ or $c^{-1}$) higher than
for the medium with nondeformable particles (see
Sec.~\ref{sec:theor_model}). Contrastingly to the case of
nondeformable particles, the latter circumstance enables us to
consider the vibration parameter $\tilde{R}_v$ to be finite for
bubbly fluids. However, for low vibration frequencies, $\Omega \gg
1$, this formalism is no longer valid. To obtain any nontrivial
effects, it becomes necessary to consider the vibration parameter
$\tilde{R}_v$ as asymptotically large, when the vibration
mechanisms studied in Sec.~\ref{sec:theor_model} become important
and cannot be neglected any more. This situation is of special
consideration and is studied in the next section.

\subsection{Low frequency approximation. Transition to a suspension of nondeformable
particles} \label{subsec:low_freq}

The case of the low (with respect to the frequency of natural
oscillation of a bubble) vibration frequency, when $\Omega \gg 1$,
should be treated separately. In this particular situation, it is
convenient to choose the value $a\omega l/\Omega^2$ as a unit for
the velocity of the radial motion, the units for the other
quantities are kept the same. As a result, the pulsation equations
read
\begin{eqnarray}
i(1-c)\hat{\bf V}=-\nabla\hat{q}, \quad \hat{\bf V}_d=3\hat{\bf
V}, \label{BF:LF:momentum_puls} \\
{\rm div}(1+2c)\hat{\bf V}=3\,\Omega_l^2c\hat{b}, \quad
\hat{q}=i\hat{b}. \label{BF:LF:mass_puls}
\end{eqnarray}
\noindent Here $\Omega_l=\Omega/l$ is the ratio of the two
asymptotic parameters. At large values of $\Omega_l$
compressibility of the bubbles becomes insignificant. We note,
that a similar result has been found out in
Ref.~\onlinecite{nigmatulin-91}: compressibility of a bubble can
be neglected only provided that the dimensionless frequency of the
radial oscillation of a bubble is large compared to $l$.

We represent the solution of the pulsation problem
(\ref{BF:LF:momentum_puls}), (\ref{BF:LF:mass_puls}) in the form
of series in the small concentration of bubbles
\begin{equation} \label{}
\hat{\bf V}=\hat{\bf V}_0+\hat{\bf V}_1+\ldots, \quad
\hat{q}=\hat{q}_0+\hat{q}_1+\ldots, \quad
\hat{b}=\hat{b}_0+\hat{b}_1+\ldots, \nonumber
\end{equation}
\noindent where $\hat{\bf V}_1$, $\hat{q}_1$, $\hat{b}_1\sim c$.

To the zero order the pulsation problem reads
\begin{eqnarray}
i\hat{\bf V}_0=-\nabla \hat{q}_0, \quad {\rm div}\hat{\bf V}_0 & =
& 0, \quad \hat{q}_0=i\hat{b}_0, \nonumber \\
{\bf n}\cdot\hat{{\bf V}}_0|_{\Gamma} & = & j_n, \nonumber
\end{eqnarray}
\noindent which admits an obvious solution
\begin{equation} \label{BF:LF:V0}
\hat{{\bf V}}_0={\bf j}, \quad \hat{b}_0=-(z-z_c).
\end{equation}
In the same way as in Sec.~\ref{subsec:BF:pulsations}, the
constant $z_c$ is obtained from the solvability condition in the
next order and takes the form (\ref{BF:z_c}).

As in Sec.~\ref{sec:theor_model}, it is necessary to take into
account a correction $\hat{{\bf V}}_1$ to the pulsation velocity.
For further purposes only the vorticity of this field turns out to
be important. Thus, from (\ref{BF:LF:momentum_puls}) we obtain
\begin{equation}
\nabla \times \hat{{\bf V}}_1=\nabla c \times {\bf j}.
\end{equation}

Equation (\ref{BF:conc_puls}) for the pulsations of concentration
transforms into
\begin{equation}
\frac{\partial \phi}{\partial
\tau}=3\frac{a}{L}\left(\Omega_l^{-2}bc - {\bf j}\cdot \nabla c
\right).
\end{equation}
\noindent Here, in contrast to Eq.~(\ref{BF:conc_puls}), the terms
in the right-hand side are of the same order and therefore have to
be retained.

Averaging of the equations is performed in the same way as in
Secs.~\ref{subsec:aver_motion}, \ref{subsec:BF:aver_motion}. As a
result we arrive at a set of the averaged equations
\begin{eqnarray}
\frac{\partial {\bf u}}{\partial t} + {\bf u}\cdot\nabla{\bf u} &
= & -\nabla \Pi + \nabla^2{\bf u}\nonumber \\
&&+{\rm R}_l\left({\bf j}{\bf j}\cdot\nabla\Phi+3A\Phi\nabla
\hat{b}^2 \right), \label{BF:LF:eq_u} \\
{\rm div}\,{\bf u} & = & 0, \quad {\bf u}_d={\bf u}+{\rm R}_l A
\nabla \hat{b}^2, \label{BF:LF:eq_u_d} \\
\frac{\partial \Phi}{\partial t} & + & {\rm div}\, \Phi{\bf u}_d =
0, \label{BF:LF:eq_c}
\end{eqnarray}
\noindent where $A=1/(4\Omega_l^2)$, ${\rm
R}_l=l^{-2}\tilde{R}_v$.

In the limit of low vibration frequency, when $A \to 0$ ($\Omega_l
\gg 1$), the model reduces to the case of nondeformable bubbles:
equations (\ref{BF:LF:eq_u})-(\ref{BF:LF:eq_c}) yield the same
model as has been developed in Sec.~\ref{sec:theor_model} [cf.
Eqs.~(\ref{aveq_vel}), (\ref{aveq_conc})]. Indeed, for the case of
bubbly fluid, $\delta \ll 1$, and inviscid pulsations, $K \gg 1$,
we have $B=1$  [see (\ref{B_inf}), (\ref{B_bubbly})] and the
vibration force $R_l\,{\bf j}{\bf j}\cdot\nabla\Phi=R_v\,{\bf
j}{\bf j}\cdot\nabla c$.

In the opposite limiting case, $A \to \infty$, the compressibility
of bubbles becomes of crucial importance; the set of equations
transforms to Eqs.~(\ref{BF:u_d})-(\ref{BF:eq_incompr}). In this
case, the solutions of the pulsation problem (\ref{BF:LF:V0}) and
(\ref{BF:V}) coincide (if measured in the same units).

\section{Evolution of a bubbly fluid in a layer} \label{sec:BF:layer}
\subsection{Statement of the problem}

A rather good understanding of vibration action on the dynamics of
a bubbly fluid can be obtained from a one-dimensional
consideration. Let us apply
Eqs.~(\ref{BF:u_d})-(\ref{BF:eq_incompr}) to study the vibration
dynamics of this medium in an infinite plane layer $-1 < z < 1$,
confined by solid walls, the vibration axis is transversal to the
boundaries. The quiescent state with the uniform distribution of
bubbles is chosen as the initial one. Symmetry of the equations,
boundary conditions, and the initial state enables us to treat the
problem in a half of the layer, where the functions $\Phi$ and
$\psi$ are even and odd, respectively. The fluid velocity vanishes
due to the incompressibility condition (\ref{BF:eq_incompr}) and
the no-slip boundary conditions at the walls. The concentration
and the potential of the pulsation velocity are defined by the
following boundary value problem:
\begin{eqnarray}
\psi^{\prime\prime}&+&\frac{3\Phi}{\Omega^2-1}\psi  =  0, \label{BF:L:gov_eq1} \\
\frac{\partial \Phi}{\partial t}&+&\left(u_d \Phi\right)^{\prime}
= 0, \quad u_d=2Q\psi\psi^{\prime}, \label{BF:L:gov_eq2} \\
z & = & 0: \; \psi=0, \, \Phi^{\prime}=0; \quad z=1: \;
\psi^{\prime}=1, \label{BF:L:bc}
\end{eqnarray}
\noindent where the prime is used to denote a derivative with
respect to $z$.

It is obvious, that the variation of the absolute value of $Q$
results only in changing of the time scale, and without loss of
generality, we set further $Q=\pm 1$ (the upper sign corresponds
to $\Omega>1$, and the lower one -- otherwise). We can also
renormalize the initial concentration in such a way that
$\Phi(z,t=0)=1$; this leads only to rescaling of the frequency
$\Omega$.

We can easily make sure that there exists no quasisteady state,
when the fluid and the bubbles are quiescent on average. Indeed,
for the quasisteady state it immediately follows from
Eqs.~(\ref{BF:L:gov_eq2})
\begin{equation}
\left( \Phi_0 \psi_0 \psi_0^{\prime}\right)^{\prime}=0,
\end{equation}
\noindent i.e., $\Phi_0={\rm const}/(\psi_0\psi_0^{\prime})$. As
earlier, the subscript ``0'' stands for indication of the
quasisteady state solution.

On the other hand, Eq.~(\ref{BF:L:gov_eq1}) then transforms to
\begin{equation}
\psi_0^{\prime\prime}+\frac{3\,{\rm
const}}{\psi_0^{\prime}(\Omega^2-1)}=0.
\end{equation}
\noindent Integrating this equation and taking into account that
$\psi_0^{\prime}$ is even, we conclude that ${\rm const}=0$ and
hence $\Phi_0=0$. Note, that this conclusion becomes invalid at
the points where $\psi_0\psi_0^{\prime}=0$.

\subsection{Initial stage of evolution}

We now turn to investigation of the initial stage of evolution,
when an analytical solution can be obtained. We assume, that
$\Phi=1+\Phi_1(z,t)$, where $|\Phi_1| \ll 1$.

Neglecting the small perturbation $\Phi_1$ in the Schr\"odinger
Eq.~(\ref{BF:L:gov_eq1}), we have
\begin{equation} \label{BF:L:inevol_psi}
\psi_0^{\prime\prime}+\frac{3}{\Omega^2-1}\psi_0=0.
\end{equation}

First, consider the case $\Omega>1$, when the vibration frequency
is lower than that of the natural oscillation of a bubble. We
introduce
\begin{equation}
\alpha^2=\frac{3}{\Omega^2-1}.
\end{equation}
\noindent and note that this expression for the wave number
$\alpha$ is in agreement with the known disperse relation for
waves in a bubbly medium (see, for example,
Ref.~{\onlinecite{caflisch-etal-85, carstensen-foldy-47}).

The solution to Eq.~(\ref{BF:L:inevol_psi}) with the
impermeability condition at the boundaries is as follows
\begin{equation} \label{BF:psi_LF}
\psi=\frac{\sin \alpha z}{\alpha \cos \alpha}
\end{equation}
The resonant frequencies correspond to the divergency of
(\ref{BF:psi_LF}) and are defined by the expression
\begin{equation} \label{BF:L:res_freq}
\Omega_n^2=1+\frac{12}{\pi^2(2n+1)^2}, \quad n=0,1,\ldots.
\end{equation}
\noindent Note, that at these frequencies it is necessary to
account for dissipation in the pulsation equations.

Taking into account the result (\ref{BF:psi_LF}), we obtain the
averaged velocity of the bubbles and the correction to the
concentration
\begin{equation}
u_d=\frac{\sin 2 \alpha z}{\alpha \cos^2 \alpha}, \quad
\Phi_1=-\frac{2 \cos 2 \alpha z}{\cos^2 \alpha}\,t+ {\rm const}.
\end{equation}

We see that the most rapid decrease of the concentration of
bubbles occurs in the vicinities of the points $z_n=\pi n/\alpha$,
$n \in Z$ -- the bubbles tend to leave the nodes of the pulsation
pressure; and vice versa, the concentration grows in the antinodes
of the pressure. This phenomenon is referred to as the primary
Bjerknes effect.\cite{bjerknes-06, feng-leal-97} However, we
stress, that in our case, the bubbles are not just advected by the
external nonuniform pulsation field (as in the conventional
Bjerknes effect). The presence of the bubbles causes this
nonuniformity.

Let us proceed to the opposite range of frequencies, $\Omega<1$.
In this case, the frequency of external action is beyond the
passband for the bubbly fluid, i.e., the waves decay deep into the
layer:
\begin{equation} \label{BF:sol_HF}
\psi=\frac{\sinh \alpha z}{\alpha \cosh \alpha}, \quad
\alpha^2=\frac{3}{1-\Omega^2}, \quad u_d=\frac{\sinh 2 \alpha
z}{\cosh^2 \alpha},
\end{equation}
\noindent Further, for the perturbation of the concentration we
find
\begin{equation}
\Phi_1=\frac{2 \cosh 2 \alpha z}{\cosh^2 \alpha}\,t+ {\rm const}.
\end{equation}

Thus, the bubbles migrate away from the boundaries of the layer to
its center. As it could be expected, at a frequency larger than
the resonant frequency of a bubble, the bubbles accumulate in the
node of the pressure, at the center of the layer.

\subsection{Finite time evolution. Numerical results}

At finite times the set of equations
(\ref{BF:u_d})-(\ref{BF:eq_incompr}), describing one-dimensional
dynamics of a bubbly fluid in the layer, was studied numerically.
The equation for transfer of the bubbles was integrated by means
of the method of characteristics. The equation for the
characteristics has an obvious form
\begin{equation}
\frac{dz}{dt}=u_d(z,t).
\end{equation}
\noindent At the characteristics the following relation is
fulfilled
\begin{equation}
\frac{d\Phi}{dt}=-u_d^{\prime}\Phi.
\end{equation}
\begin{figure}[!b]
\vspace{-2.5mm}
\includegraphics[width=5.0cm]{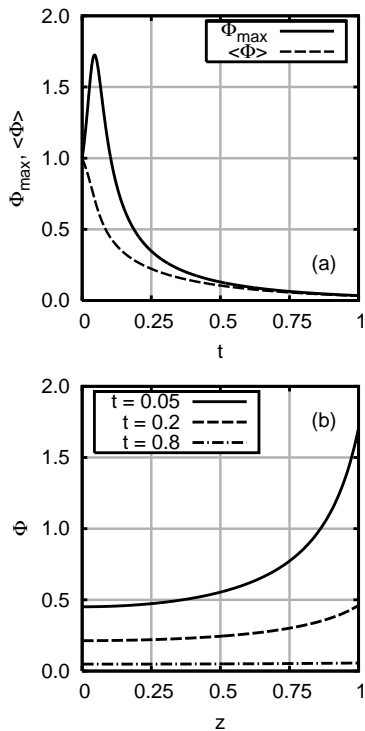}
\caption{The maximal $\Phi_{max}$ and the averaged
$\left<\Phi\right>$ values of concentration as functions of time
(a); the concentration profiles at different times, $Q = 1,
\Omega^2 = 3$ (b).} \label{fig:4}
\end{figure}

The fields of the concentration $\Phi$ and potential $\psi$ are
defined in the nodes of the uniformly spaced grid; we used up to
2000 grid nodes. In the case when $u_d(1)<0$, there appears a
front of the concentration $z=z_s(t)$ such that $\Phi \ne 0$ only
in the domain with $z<z_s$. For this particular situation, the
grid nodes were interposed only in the domain occupied by the
bubbly fluid, i.e., for $z<z_s(t)$; in the opposite part of the
layer, the equations admit an obvious solution: $\Phi=0$,
$\psi=\psi_0+z$. At the point $z=z_s$ the functions $\psi$,
$\psi^{\prime}$ are continuous, so that the value of the constant
$\psi_0$ is insignificant. In the opposite case, $u_d(1)>0$, a
part of bubbles settles on the boundaries: the averaged
concentration $\left<\Phi\right>=\int_{0}^{1}\Phi \, dz$ decreases
with time.

Results of the numerical simulation are presented in
Figs.~\ref{fig:4}-\ref{fig:6}. If $\Omega>\Omega_0=1.489$ [for the
uniform distribution of the bubbles, see relation
(\ref{BF:L:res_freq})], the bubbles gradually leave the bulk of
the layer, settling on the boundaries, Fig.~\ref{fig:4}. At first,
the maximal value $\Phi_m$ of the concentration grows and after a
while starts to decrease; the averaged concentration $\left< \Phi
\right>$ monotonically decreases with time.
\begin{figure}[!t]
\vspace{-2.5mm}
\includegraphics[width=5.5cm]{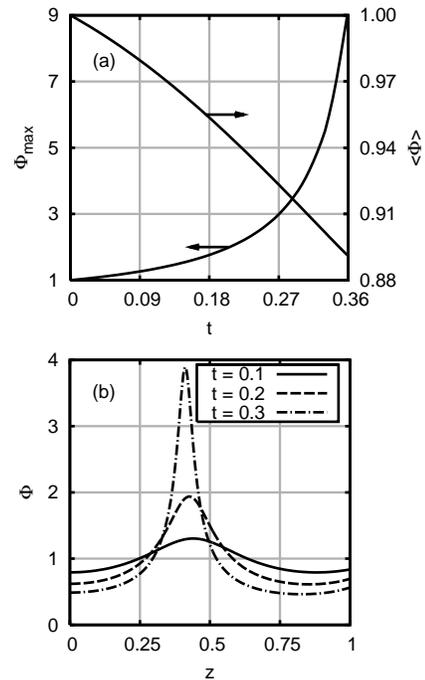}
\caption{The maximal $\Phi_{max}$ and the averaged
$\left<\Phi\right>$ values of concentration as functions of time
(a); the concentration profiles at different times, $Q = 1,
\Omega^2 = 1.25$ (b).} \vspace{-2mm} \label{fig:5}
\end{figure}

In the case of $\Omega_1<\Omega<\Omega_0$, the bubbles accumulate
in a certain domain inside the layer. The concentration of the
bubbles rises abruptly near the point $z_s$ and reaches infinite
values for a finite time $t=t_0$ -- there develops a peaking
regime. The maximal concentration grows as $(t-t_0)^{-1}$, i.e., a
bubble screen develops. The potential itself remains finite, while
its second derivative tends to infinity near the point $z_s$.
%
% Later there develops a bubble screen -- a $\delta$-like
% distribution of the bubbles, at which the derivative of the
% potential becomes discontinuous (see, for example,
% Ref.~\onlinecite{flugge-94}):
% %
% \begin{equation}
% \Phi=\Phi_0\delta(z-z_s)+{\rm n.s.p.}, \quad
% \psi^{\prime}(z_s+\epsilon)-\psi^{\prime}(z_s-\epsilon)=-\frac{3\Phi_0}{\Omega^2-1}\psi(z_s).
% \end{equation}
% %
% \noindent Here, ${\rm n.s.p.}$ is the nonsingular part of the
% concentration; $\Phi_0(t)$ and $z_s(t)$ are the power and the
% coordinate of a given bubble screen, respectively.
This scenario is demonstrated for $\Omega^2=1.25$ in
Fig.~\ref{fig:5}. The averaged concentration of the bubbles
decreases with time. As the eigenfrequency $\Omega$ gets closer to
$1$, the number of bubble screens increases.

In the case $\Omega<1$, the bubbles accumulate in the center of
the layer -- a node of the pulsation pressure. Evolution of the
maximal concentration, the coordinate $z_s$ of a front, i.e., the
interface bubbly fluid--pure fluid, and the profiles of the
concentration are depicted in Fig.~\ref{fig:6}. As was mentioned
above, the bubbles located near the front possess the maximal
velocities [cf. (\ref{BF:sol_HF})], consequently the concentration
grows very fast at the front and does not practically vary in the
stagnant core in the center of the layer.

% In the limiting case of weak compressibility, $\Omega \gg 1$, the
% one-dimensional solution to
% Eqs.~(\ref{BF:LF:eq_u})-(\ref{BF:LF:eq_c}) is straightforward
% %
% \begin{equation}
% u=0, \quad u_d=2\,{\rm R}_l Az \; (z_s=0), \quad
% \Phi=\Phi_0\exp(-2\,{\rm R}_l A t). \nonumber
% \end{equation}
% %
% \noindent Thus, under vibration the fluid gets free of bubbles,
% the distribution of the bubbles {\bf remains/keeps} uniform.
%
\begin{figure}[!h]
\vspace{-2.5mm}
\includegraphics[width=5.5cm]{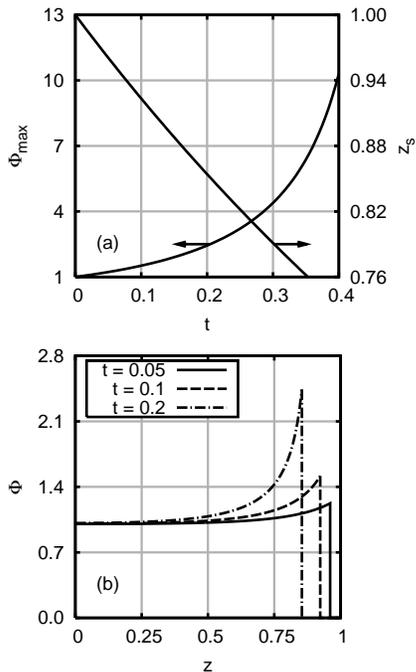}
\caption{The maximal concentration $\Phi_{max}$ and the coordinate
of the front $z_s$ as functions of time (a); the concentration
profiles at different times, $Q = -1, \Omega^2 = 0.5$ (b).}
\vspace{-0mm} \label{fig:6}
\end{figure}

\section{Conclusions} \label{sec:conclus}

The averaged dynamics of various two-phase systems in a
high-frequency vibration field has been theoretically studied. The
continuum approach was applied to describe such systems as solid
particle suspensions, emulsions, bubbly fluids, when the volume
concentration of the disperse phase is small and gravity is
insignificant. The dynamics of a monodisperse system was
considered by means of the averaging method, when the fast
pulsation and slow averaged motion can be treated separately. The
vibration dynamics of suspensions of both nondeformable and
deformable particles, when the compressibility of the disperse
phase becomes important, has been investigated.

An averaged model for nondeformable particles has been obtained.
It is shown, that accounting for the Stokes drag and unsteady
forces such as the Basset history term, the second memory force
and the added mass force in the pulsation motion, is of crucial
importance for correct description of the averaged dynamics. The
averaged equations have been obtained and simplified in the
framework of the single-fluid approximation. It is proven that the
particles can be treated as frozen into the averaged fluid flow.
As a result the averaged dynamics of a suspension is described by
the equations of momentum and mass conservation for the carrier
fluid and a transfer equation for the concentration of particles.
The action of vibration results in the appearance of a vibration
force in the equation of the fluid motion. This force is
nonvanishing only for the nonuniform distribution of particles,
its direction coincides with the vibration axis. We note, that the
developed model is expected to be worth applying to clarify recent
experiments on blood flow resistance under vibration action.
\cite{shin-etal-03}

The developed model has been applied to study the behavior of a
two-phase medium in an infinite plane layer subjected to
transversal vibrations. It is demonstrated that there is a set of
possible nonuniform quasiequilibrium distributions of particles,
when the averaged flow vanishes but the pulsation velocities do
not. The stability of a linear distribution of particles has been
investigated. Analytical and numerical analysis manifests that the
quasiequilibrium state is unstable for any intensities of
vibration. A nonuniformity of particle distributions causes the
vibration force that generates the averaged motion in fluid.
Because of this flow, the initial distribution of particles is
deformed and the state becomes unstable.

As a system with deformable particles, the case of bubbly fluids
in vibration field has been analyzed separately. An averaged model
accounting for the compressibility of the bubbles has been
developed. The pulsations were assumed to be inviscid, and a set
of averaged equations has been obtained in the single-fluid
approximation. The averaged velocity of bubbles differs from the
fluid velocity, this difference is caused by vibration and
proportional to its intensity. It is shown that in contrast to the
case of nondeformable particles, the impact of vibration on the
system with deformable particles is significantly stronger. Even
for uniform distribution of particles the vibration force is
nonzero.

An intermediate case of low vibration frequencies, when the ratio
$\Omega$ of eigenfrequency of radial bubble oscillations to the
the frequency of vibration is high, has been considered. A
criterion when the compressibility of bubbles can be neglected has
been figured out: $\Omega \gg L/R$, where $L$ is the length scale
of the flow and $R$ is the time-averaged radius of the bubble. In
this case the intermediate model reduces to the model for
nondeformable particles. In the opposite limit, $\Omega \ll L/R$,
the intermediate model transforms to the discussed model where
compressibility of bubbles becomes of crucial importance.

The dynamics of bubbly fluid in an infinite plane layer under the
action of transversal vibration has been analyzed. The quiescent
state with the uniform distribution of bubbles is chosen as the
initial one. It turns out that for $\Omega > 1.489$ the bubbles
migrate to the boundaries of the layer; for $1<\Omega<1.489$ the
bubble screens appear; for $\Omega<1$ bubbles accumulate in the
center of the layer. We have demonstrated that the behavior of
bubbly fluid is analogous to the primary Bjerknes effect: for
$\Omega > 1$ the bubbles leave the nodes and accumulate in the
antinodes of the pressure wave, while for $\Omega < 1$ the bubbles
migrate to the nodes. However, in our case, the bubbles are not
only advected by the external nonuniform field, as in the
conventional Bjerknes effect, but also cause this nonuniformity.

%\acknowledgments
\section{Acknowledgments}

The research was partially supported by RFBR (Grant
No.~04-01-00422) and CRDF (Grant No.~PE-009-0), which are
gratefully acknowledged. A.S. thanks the German Science Foundation
(DFG, SPP 1164); S.S. is thankful to DAAD and Russian Ministry of
Education and Science (Russian-German Mikhail Lomonosov Program).

\appendix*
\section{}

Here we analyze relation (\ref{param_B}) as a function of the
parameters $\delta$, $\kappa$ and $K$. First, we give the
expressions for the limiting vibration regimes of viscous ($K \ll
1$) and inviscid ($K \gg 1$) pulsation. Next, we discuss the
dependence in the whole range of values of $K$ and outline
features for some typical media.

In the limit of viscous pulsation, $K \ll 1$, relation
(\ref{param_B}) reduces to the following:
\begin{equation}
B_{0}=\frac{(\delta-1)^2 K^3}{9\sqrt{2}}.\label{B_0}
\end{equation}
\noindent Note, in this limit, it is not enough to take into
account only the Stokes force, which would result in vanishing $B$
and therefore in no vibration force. Physically, this means that
the particles get frozen not only in the averaged flow, but in the
pulsation fluid flow as well, and therefore no averaged effects
are possible. To account for any nontrivial dynamics one has to
retain small (with respect to the Stokes drag) terms due to memory
forces, which ensures relative pulsation motion of phases. Another
interesting point is that this result does not depend on $\kappa$.

In the opposite limit of inviscid pulsation, $K \gg 1$, the
leading contribution to $B$ is made by the added mass force. The
result is sensitive solely to the relative density of phases:
\begin{equation}
B_{\infty}=\left(\frac{\delta-1}{1+2\delta}\right)^2.\label{B_inf}
\end{equation}

A typical case of a bubbly fluid ($\kappa \ll 1$, $\delta \ll 1$)
is governed by
\begin{equation} \label{B_bubbly}
B_{b}=\frac{2K^3(18\sqrt{2}+9K+3\sqrt{2}K^2+K^3)}{(18\sqrt{2}+18K-K^3)^2+K^2(18+3\sqrt{2}K+K^2)^2}.
\end{equation}
\noindent The result is caused by the Stokes drag, the second
memory force, and the added mass force.

For the case of solid particles ($\kappa \to \infty$), the
dependence (\ref{param_B}) transforms to:
\begin{equation} \label{B_solid}
B_{s}=\frac{(\delta-1)^2\sqrt{2}K^3(9+\sqrt{2}K)}{81(\sqrt{2}+K)^2+K^2[9+\sqrt{2}(1+2\delta)]^2},
\end{equation}
\noindent where the Stokes drag, the Basset and the added mass
forces make finite contribution.

To demonstrate the dependence in more general situations, we
tabulate (\ref{param_B}) as a function of $K$ for different values
of $\delta$ and $\kappa$. Typical results are presented in
Fig.~\ref{fig:7}. As it can be shown, the function $B(K)$ is
positive for all $K$; it is either monotonic or has a maximum. In
the limits of $K \ll 1$ and $K \gg 1$ the results are independent
of $\kappa$ and agree with the expressions (\ref{B_0}) and
(\ref{B_inf}), respectively. As it can be expected, the effect of
different values of $\kappa$ occurs at intermediate values of $K$.
\begin{figure}[!h]
\vspace{-2.5mm}
\includegraphics[width=4.7cm]{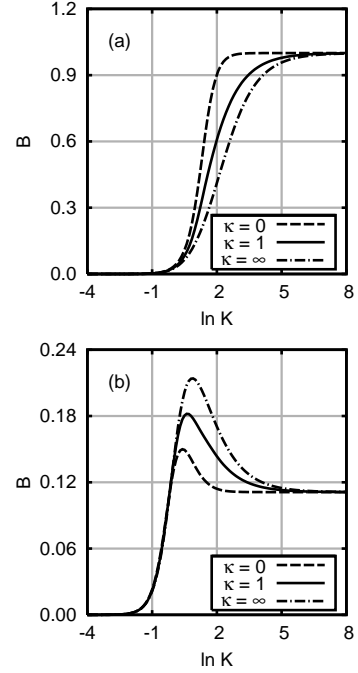}
\caption{The dependence of $B$ against the parameter $K$ for
$\delta=0$ (a) and $\delta=4$ (b).} \label{fig:7}
\end{figure}

Figure~\ref{fig:7}(a) represents the case of bubbly fluids
($\delta \ll 1$); the curve with $\kappa=0$ corresponds to
relation (\ref{B_bubbly}). The curves go out from the point $B=0$
at $K=0$, start to grow at small $K$ as $K^3/(9\sqrt{2})$ [see
relation (\ref{B_0})], and monotonically approach the value
$B_{\infty}=1$, as follows from (\ref{B_inf}). For $\delta \le
1/2$, qualitative form of the curves keep the same as in
Fig.~\ref{fig:7}(a), the value of $B_{\infty}$ changes according
to (\ref{B_inf}).

For $\delta > 1/2$ the dependence $B(K)$ is no longer monotonic:
there appears a maximum $B_{m}$ at some value $K_m$ [see
Fig.~\ref{fig:7}(b)]. The curve with $\kappa \gg 1$ corresponds to
solid particles and is described by formula (\ref{B_solid}); such
a curve in Fig.~\ref{fig:7}(b) is plotted for a suspension ``sand
in water'' ($\delta=4$, $\kappa \gg 1$). We note, that for higher
relative densities $\delta$ the dependence $B(K)$ does not change
qualitatively. In the limiting case of heavy particles suspended
in a gaseous medium (solid particle suspension in air, aerosol),
$\delta \gg 1$, the maximum value $B_m$ grows and shifts to lower
values of $K$; the asymptotic behavior reads $B_m \propto
\sqrt{\delta}$, $K_m \propto 1/\sqrt{\delta}$. At high values of
$K$ parameter $B$ gradually approaches the limiting value
$B_{\infty}=1/4$.

\end{document}